\documentclass[letter,onecolumn, pre,showkeys]{revtex4}

\usepackage{nicefrac}                 
\usepackage{latexsym,amsmath,amssymb,amsfonts} 
                                
\usepackage{amsthm}             
\usepackage{subeqnarray}        
                                
\usepackage{float}              
\usepackage{graphicx}           

\newcommand{\ud}{\ensuremath{\mathrm{d}}}           
\newcommand{\sech}{\ensuremath{\mathrm{sech}}}
\newcommand{\csch}{\ensuremath{\mathrm{csch}}}
\newcommand{\cn}{\ensuremath{\mathrm{cn}}}
\newcommand{\sn}{\ensuremath{\mathrm{sn}}}
\newcommand{\dn}{\ensuremath{\mathrm{dn}}}

\begin{document}

\title{Tilted excitation implies odd periodic resonances}                                  

\author{G.I. Depetri}
\email{gdepetri@if.usp.br}
\affiliation{Instituto de F\'isica ``Gleb Wataghin'', Universidade
  Estadual de Campinas, 13083-859 Campinas, SP, Brazil} 
\affiliation{Instituto de F\'isica da Universidade de S\~ao Paulo,
  05315-970 S\~ao Paulo, SP, Brazil}

\author{J.C. Sartorelli}
\email{sartorelli@if.usp.br}
\affiliation{Instituto de F\'isica da Universidade de S\~ao Paulo,
  05315-970 S\~ao Paulo, SP, Brazil}

\author{B. Marin}
\affiliation{Instituto de F\'isica da Universidade de S\~ao Paulo,
  05315-970 S\~ao Paulo, SP, Brazil}
\affiliation{Department of Neuroscience, Physiology and Pharmacology,
  University College London, Gower Street, London, WC1E 6BT, UK}

\author{M.S. Baptista}
\affiliation{Institute for Complex Systems and Mathematical Biology,
  SUPA, University of Aberdeen, Aberdeen, AB24 3UE, UK}

\begin{abstract}
Our aim is to unveil how resonances of parametric systems are affected
when symmetry is broken. We showed numerically and experimentally that
odd resonances indeed come about when the pendulum is excited along a
tilted direction. 
Applying the Melnikov subharmonic function, we not only determined
analytically the loci of saddle-node bifurcations delimiting resonance
regions in parameter space, but also explained these observations by
demonstrating that, under the Melnikov method point of view, odd
resonances arise due to an extra torque that appears in the asymmetric
case.      
\end{abstract}


\maketitle

\section{Introduction}
Resonance is an ubiquitous phenomenon. It provides an ideal landscape
for efficient functioning of many natural and technological systems,
such as the amplification of speech in body cavities \cite{speech},
the motion of a child on a swing \cite{swing}, imaging by magnetic
resonance and electronic signal amplification \cite{par-amplifiers,
  optical, josephson}. Parametric resonance, in particular, arises
from time-dependent modulation of system parameters, and has important
implications in phenomena ranging from boat capsizing in naval
engineering \cite{boris-naval}, encoding of tactile information in
rodent whisking behaviour \cite{boris-neuro} and energy harvesting
from mechanical oscillatory motion \cite{waves-1}. 
The parametric pendulum, described by  
\begin{equation} \label{vertical}
  \ddot{\theta} + \beta \dot{\theta} + (1 + P \cos \Omega_p t) \sin
  \theta = 0, 
\end{equation}
is a paradigmatic model in nonlinear dynamics
\cite{note-mathieu,mathieu, butikov-subharm}. 
It consists of a planar simple pendulum whose pivot oscillates
harmonically along the vertical direction with amplitude $P$ and
frequency $\Omega_p$, and $\beta$ is the friction parameter. This
system has attracted great attention \cite{bif-control,
  nearly-parametric, rega-competing, analytical-solutions,
  generalized, clifford-bishop-control}. It presents a wide range of 
dynamical behaviour, such as: the stabilization of the hilltop 
saddle \cite{invertido, smith-invertido-exp, mult-nodding,
  clifford-bishop-inverted, bishop-sudor-inverted, butikov-inverted};
the occurrence of chaotic behaviour 
\cite{koch-leven-ch-behaviour, koch-leven-evidence,
  koch-leven-experiments, clifford-bishop-zones, poloneses-1,
  poloneses-2}; the observation of period-doubling cascades 
\cite{cascade, periodo-impar} and the existence of resonance
regions \cite{butikov-subharm, verhulst}. Moreover, it can be used as
qualitative analogue for more complex systems \cite{periodo-impar,
  analogia, chemical}. Indeed, mechanical analogues of physical
systems provide a direct visualization of motion allowing an intuitive
understanding of the system being studied, as it is done for the
analysis of power grid \cite{doerfler} and applications of
telecommunications \cite{americano}. Rotating motion in the parametric
pendulum has also been widely considered in the literature
\cite{kapitaniak}, due to the possibility of energy harvesting from
sea waves, which would consist in transforming the vertical motion of
sea waves in rotating motion of systems composed by parametric pendula
\cite{waves-1, clifford-bishop-rotating, xuxu-rotating, 
  rotating-lenci, rotating-lenci-2, xuxu-rotating}.   

The study of stable periodic orbits of Eq. \eqref{vertical} has a long
academic history, in particular regarding those whose period is an
even or odd multiple of the excitation period
\cite{note-periodo-n}. Previous works show the existence of both even 
\cite{per-doubling, bryant-3} and odd \cite{periodo-impar, period-3,
  clifford-bishop-tangencies} oscillations in the frequency region
$\Omega_p<2$. Odd resonances however are very rarely reported,
supposedly due to its non-typical nature (only observed for narrow
ranges of parameter values, according to Ref. \cite{period-3}).  
To describe theoretically the mechanisms that could lead to
subharmonic solutions, Koch and Leven \cite{koch-leven} applied the
Melnikov theorem for subharmonic bifurcations \cite{guck-holmes,
  wiggins} to this system. They succeeded in calculating parameter
ranges for the existence of even oscillations, but concluded that the
Melnikov theorem could not be applied to unravel parameter ranges for
odd oscillations. This is also true when Melnikov method is applied to
other similar parametric systems \cite{note-parametric, nonharmonic,
  belyakov}.  
In Ref. \cite{torque}, it is shown that for nonlinear perturbed
systems, even oscillations are due to parametric excitation along the
gravitational field, while odd oscillations can exist as consequence
of external torques. Looking at Eq. (\ref{vertical}), one notices that 
this system is symmetric with respect to the transformation 
$\theta \rightarrow - \theta$. External torques are absent. However,
breaking the symmetry of Eq. (\ref{vertical}) by a tilt in the
pendulum pivot motion should introduce an additional torque in the
equations of motion, as we shall see. A question then arises: does
symmetry breaking affect somehow the behavior of odd oscillations? 

We have demonstrated numerically and experimentally, for the tilted
parametric pendulum, the existence of odd resonances in the frequency
region $\Omega_p>2$. Also, applying the Melnikov subharmonic
function, we obtained analytically the loci of saddle-node
bifurcations that are the parameter thresholds for the existence of
odd stable oscillations. Surprisingly, these loci coincide with the
curves that limit resonance regions found numerically. Moreover, we 
show that according to the Melnikov method approach, whereas even
resonances are a consequence of the vertical excitation, odd ones
comes about due to the tilt in the orientation of the excitation with
respect to the vertical position. This work therefore paves the way
to a better understanding of how resonances of any type can appear in
complex oscillating systems and how they can be related to symmetry.

\section{Experimental apparatus}
\begin{figure}[htb!]
  \centering
  \includegraphics[width=7.8cm,height=6.2cm]{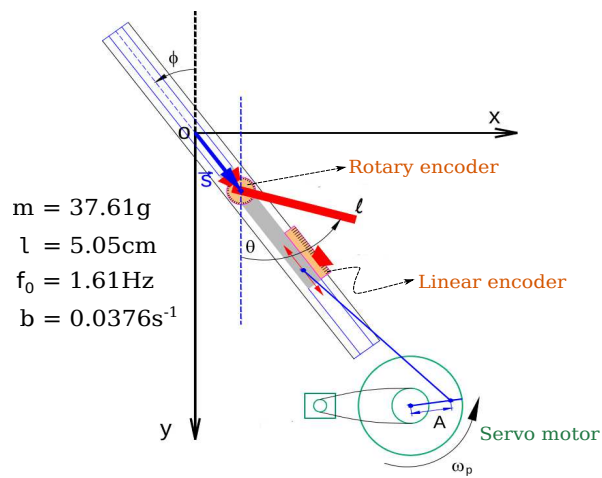}
  \caption{(Color online). Diagram of the experimental apparatus of a
    planar simple pendulum parametrically excited along an arbitrary
    direction. The pendulum pivot axis is attached to a sliding
    car that is forced to oscillate with amplitude $A$ and frequency
    $f_p$, with the help of a crank and a servo motor. We can measure 
    simultaneously the absolute value of the pivot speed $v_p$, with a
    linear encoder system in the lab reference frame, and the absolute
    value of the pendulum angular velocity $|\omega|$, with a rotary
    encoder system attached to the pendulum.}    
  \label{apparatus}
\end{figure}

A diagram of the experimental apparatus is shown in
Fig. \ref{apparatus}. The pendulum consists of a single arm with mass  
$m$ and center of mass position at a distance $l$ from the pivot
axis. The natural frequency of oscillation is $f_0$ and the friction
parameter is $b$ \cite{note-pendulum-parameters}. The angle between 
the pendulum arm and the vertical direction is $\theta$. 
The pivot of the pendulum is attached to a sliding car, which is
periodically excited according to $s=A\cos\omega_pt$, along a tilted
axis making an angle $\phi=\nicefrac{\pi}{8}$ with the vertical
direction. The rail over which the pivot oscillates is attached to
the wall, therefore $\phi$ is not a dynamic parameter.
To measure the absolute value of the pivot velocity
$v_p=|\dot{s}|$, we attached a linear optical encoder, of resolution
$\frac{2.54}{500}$ cm, to the sliding car. From the time series
$v_p(t)$, we can obtain the frequency $f_p=2\pi\omega_p$ and the
amplitude $A$ of the external excitation, which are the control
parameters of the system.   

To measure the absolute value of the angular velocity
$\omega=\dot{\theta}$ of the pendulum arm, we attached an optical
rotary encoder, of resolution $\frac{2\pi}{2500}$ rad, to the pendulum
arm, concentric to the pivot point. The light sensor of the rotary
encoder is attached to the sliding car, concentrically to the pivot,
while the sensor of the linear encoder is attached to the lab reference
frame. The induced inputs from both encoders were detected by using an 
ADC board of 16 bits at the rate of 200 Ksamples$/$s. For fixed values
of $A$, the excitation frequency $f_p$ was spanned in the forward and
backward directions with a servo motor in steps of $0.01$Hz. 

\section{Equations of motion}
\label{EOM}
The equations of motion of the undamped parametric pendulum can be
obtained via Lagrangian formulation. Adding a linear damping
term $-b \dot{\theta}$, we arrive at  
\begin{equation} \label{pendulo2d}
\begin{split}
  \dot{\theta} &= \omega, \\
  \dot{\omega} &= - \sin \theta -  \left[ P
    \cos(\Omega_p \bar{t}) \sin(\theta - \phi) + \beta \omega \right], 
\end{split}
\end{equation}
where the dot indicates derivative with respect to the
normalized time $\bar{t}=\omega_0t$, and
\begin{equation} \label{parameters}
  \Omega_p=\frac{\omega_p}{\omega_0}, \qquad 
  P=\frac{\omega_p^2A}{g}, \qquad
  \beta=\frac{b}{\omega_0},
\end{equation}
are the dimensionless parameters of the system. The period of the
perturbation is $T_p = 2\pi/\Omega_p$. Notice the symmetry
$\theta \rightarrow - \theta$ is now broken, and comparing with 
Eq. \eqref{vertical} we have an extra torque given by
$\tau=P\sin{\phi}\cos{(\Omega_p\bar{t})}\cos{\theta}$, non null for
$\phi\neq0$. 

\section{Data analysis}
\label{strobo_map}
\begin{figure}[htb!]
  \centering
  \includegraphics[width=8.5cm]{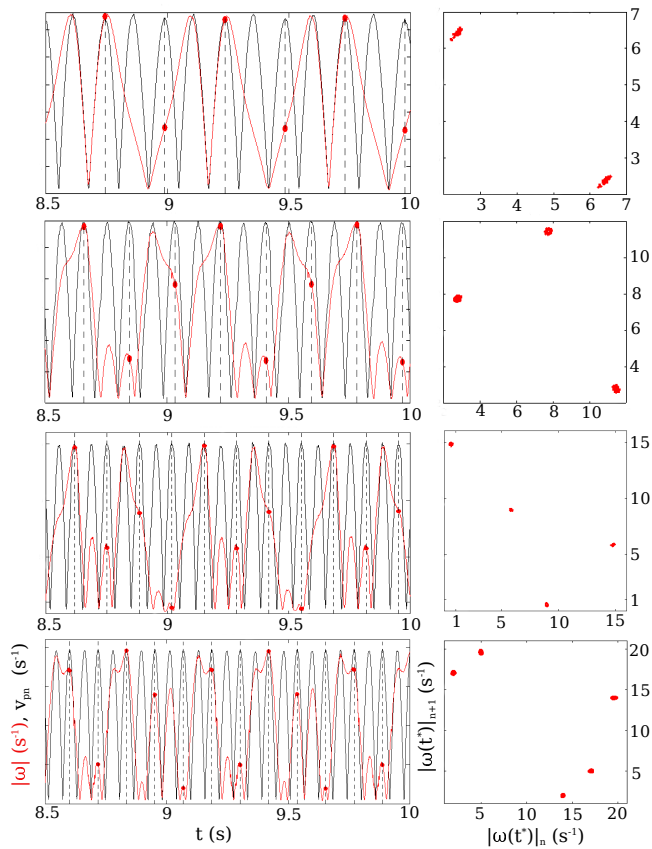}
  \caption{(Color online). On the left column, from the top to the
    bottom, we display sampled experimental time series for fixed
    amplitude $A=2.02$cm illustrating period-2 ($f_p=4.05$Hz),
    period-3 ($f_p=5.35$Hz), period-4 ($f_p=7.5$Hz), and period-5 
    ($f_p=8.55$Hz) oscillations. Full red lines represent the modulus
    of the pendulum angular velocity, $|\omega|$, and full black lines
    the time series for the modulus of perturbation velocity
    normalized with respect to the maximum value of $|\omega|$,
    $v_{pn}$. Full red circles are the values of the pendulum velocity
    each time the perturbation completes a full cycle. On the right
    column we have the corresponding stroboscopic maps for each time 
    series.}   
  \label{data}
\end{figure}

The periodicity of each trajectory, both in numerical and experimental
analyses, was computed from the stroboscopic map for the absolute
value of the angular velocity of the pendulum 
$|\omega[(k+1)T_p]| \times |\omega(kT_p)|$, $k\in\mathbb{N}$.
The stroboscopic map is obtained by observing the (absolute) values of
the pendulum velocity $|\omega(t)|$ every time the perturbation
completes a full cycle, i.e., every second maximum of $v_p$.
In Fig. \ref{data}, sampled experimental time series for period-2, 3,
4 and 5 oscillations are displayed, as well as the respective
stroboscopic maps. The number of points $m$ in the stroboscopic map is 
related to the ratio between the period of oscillation and the period 
of the excitation, being that $T/T_p = m/n$, with $m$ and $n$ co-prime 
integers. 

\section{Numerical results}
\begin{figure}[htb!]
  \centering
  \includegraphics[width=8.5cm,height=9cm]{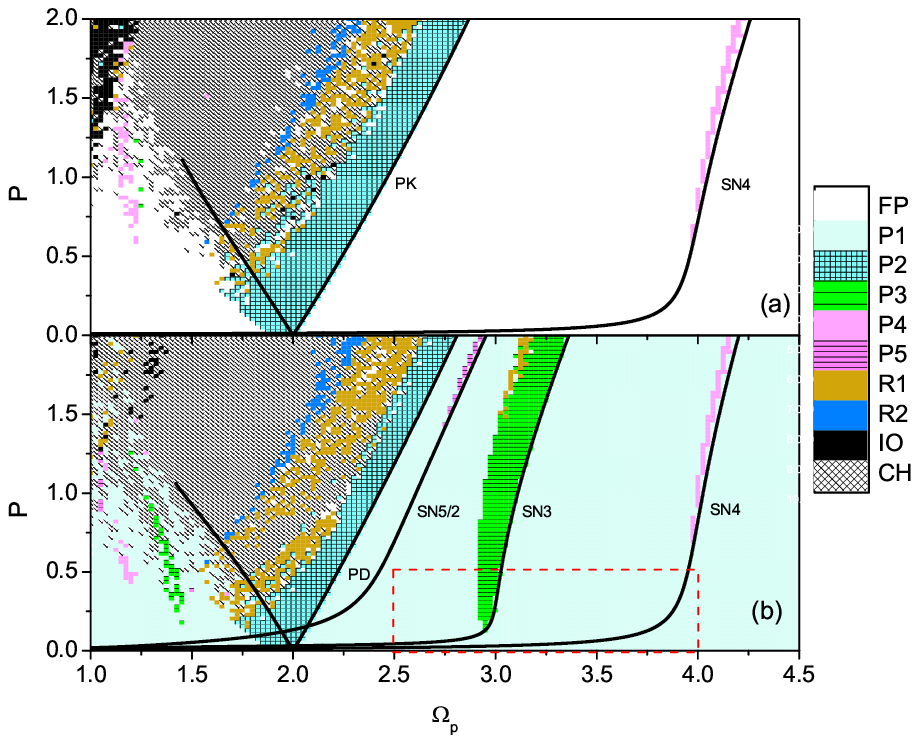}
  \caption{(Color online). Parameter spaces for: (a) vertical
    $(\phi=0)$; and (b) tilted $\left(\phi=\pi/8\right)$
    excitations. FP denotes fixed point; P$m$ and R$m$, period-m
    \cite{note-periodo-n} oscillation and rotation, respectively; IO
    periodic inverted oscillations, and CH persistent tumbling chaos
    \cite{clifford-bishop-zones}. Full lines represent the loci of
    saddle-node (SN), pitchfork (PK) and period-doubling (PD)
    bifurcations, and were obtained with the software AUTO 07p
    \cite{auto, auto-2}.}
  \label{esp-par}
\end{figure}

In Fig. \ref{esp-par}  parameter spaces for the symmetric (a) and
asymmetric (b) cases are shown. They are constructed using direct
integration. For each pair of parameters ($\Omega_p$, $P$) we 
integrate Eqs. \eqref{pendulo2d} for a time span $\bar{t}=$ 1000, with 
initial conditions $\theta_{i}=0.55$ and $\omega_{i}=0$ and, after 
discarding initial data up to $\bar{t}=$ 900 (transient behaviour), we
compute the periodicity of the stationary solution as explained in
Section \ref{strobo_map}. Rotations and oscillations of different
periods, as well as irregular motion, are observed. 

In Fig. \ref{esp-par}(a), we display the parameter space for the
vertical excitation case. Notice that off the resonance regions,
the pendulum does not oscillate, but stays at the fixed point (FP)  
$\theta=\omega=0$. From the right to the left, full line SN4 is the
loci of saddle-node bifurcations for which a period-4 stable
oscillations (P4) come about, and full line PK is the loci of
pitchfork bifurcations at which stable period-2 (P2) oscillations
appear. Then we have period-1 (R1) and period-2 (R2) rotations, and
persistent tumbling chaos (CH).  
In Fig. \ref{esp-par}(b) we have the parameter space for the
non-symmetric perturbation case. Regions P1 represent parameters
generating non resonant regions, when the pendulum oscillates with
the same period of the perturbation. From the right to the left, we
see with the full lines SN4, SN3 and SN5/2, respectively, the loci of
saddle-node bifurcations from which period-4 (P4), period-3 (P3) and
period-$\nicefrac{5}{2}$ stable oscillations come about. Full line PD
is the loci of period-doubling bifurcations in which the period-1
oscillations lose stability and stable period-2 resonant stable
oscillations (P2) appear. We observe period-1 (R1) and period-2 (R2) 
rotations, and then persistent tumbling chaos (CH). Multistability is
present for some parameter values, but since the integration has only
been performed for a single initial condition, this cannot be seen
here.
In the vertical case we observed only even resonances, and in tilted
case, both even and odd resonances. However, we must have in mind that
the parameter space depends on the choice of initial conditions. 

For the parametrically excited double pendulum, regions of
non-oscillation and the pitchfork bifurcations in the vertical case are 
substituted for period-1 oscillations and period-doubling bifurcations
when excitation is tilted \cite{sarto-1, sarto-2}. This is due to the
extra torque introduced by symmetry breaking, as seen in
Section \ref{EOM}.  

\section{Experimental results}
\begin{figure}[htb!]
  \centering
  \includegraphics[width=8.5cm]{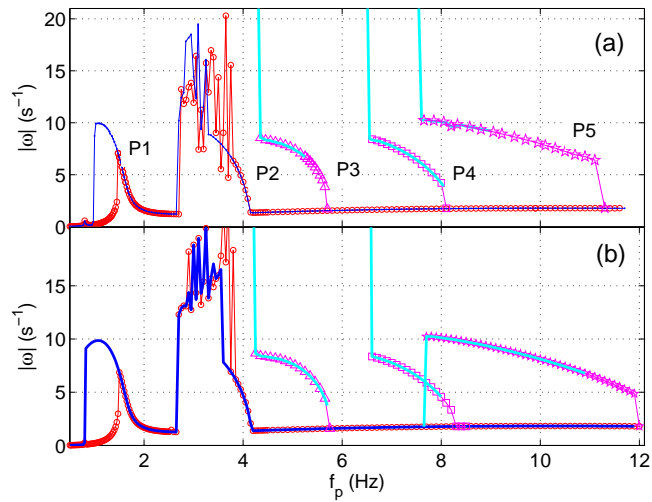}
  \caption{(Color online). Bifurcation diagrams at $A = 2.02~$cm
    obtained (a) experimentally and (b) numerically. Frequency of the
    excitation was spanned both forward (line with empty circles)
    and backwards (full line). Periodic oscillations P3, P4 and
    P5 are excited by manually setting up initial conditions of the
    pendulum, and are displayed in triangles, squares and stars,
    respectively, for forward frequency sweeping, and in full lines
    for backward sweeping.}  
  \label{experimental}
\end{figure}

We confirm experimentally the behaviors in Fig. \ref{esp-par}(b)
through bifurcation diagrams, by fixing the amplitude of excitation at 
$A=2.02~$cm and performing both forward and backward frequency
sweeps. Results are shown on Fig. \ref{experimental}(a). The occurence  
of odd primary resonances P1, P3 and P5 for large parameter ranges is
remarkable.
It is important to observe that the control parameter in the
laboratory are the amplitude $A$  and the frequency $f_p$ of the
excitation, while the parameter space is computed in terms of the
normalized parameters $P$ and $\Omega_p$. Since
$P = \frac{\omega_0^2}{g} A \Omega_p^2$, by keeping $A$ constant,
we have a parabola in the parameter space of Fig. \ref{esp-par}.
Also, all points in parameter space are generated by integrating
the equations of motion with the same initial conditions. When the
experiment is being performed, the inicial conditions are
automatically updated from point to point. To show the remarkable
agreement between numeric and experimental results, in
Fig. \ref{experimental}(b) we show the equivalent bifurcation diagram 
computed numerically. 

Along the line with empty
circles (increasing $f_p$), from the left to the right, the pendulum
starts realizing small oscillations with the same period and phase of
the excitation. At $f_p=\frac{f_0}{2}=0.8$Hz, we have a bifurcation at
which it oscillates with period $\frac{T_p}{2}$ and out of phase with
perturbation. Another bifurcation happens at around $f_p=f_0=1.61$Hz, 
causing it to oscillate with the same period of the perturbation, but
different phase. For larger frequencies, at around $f_p=2$Hz, the
pendulum slowly gets in phase with the excitation. For around
$f_p=2.8$Hz irregular motion is observed and we needed to prevent
the pendulum from rotating, because the energy gain is so large that
it can destroy the apparatus \cite{sarto-1}. Around
$f_p=2f_0=3.22$Hz the pendulum oscillates with twice the period of 
excitation, and for a slight increase in frequency it goes back to
oscillating with the same period and phase of the perturbation until
the end of data acquisition. Following the full line (decreasing
$f_p$), we observe basically the same phenomena, except for when there
is hysteresis. In this process, we observe no oscillation with period
higher than $2T_p$, but this is because in the frequency region where
we expect to observe those solutions, the pendulum is oscillating in
the regime of small oscillations, outside the basins of attraction of
these long-period attractors.     

To detect those solutions, we need to manually adjust the initial
conditions of the pendulum in the parameter region we expect to
observe them. Then, once the wanted attractor is found, we perform
forward and backward frequency sweeps.  
If we adjust $\theta_0 \approx 0.5$ in the region
$f_p \approx 3f_0=4.83~$Hz we find the period-3 resonant oscillatory
attractor P3. The triangles represent those solutions with increasing
frequency, and the full line above them, the same solutions with
decreasing frequency. Also, if we adjust $\theta_0\approx0.36$ in the
region $f_p\approx4f_0=6.44~$Hz, we find the period-4 oscillatory
attractor P4. The squares represent those solutions with increasing
frequency and, as before, the full line above them, the same solutions
with decreasing frequency. We also found period-5 oscillations we did
not foresee numerically from Fig. \ref{esp-par}, because the initial
conditions we chose were out of its basin of attraction. If we adjust
$\theta_0\approx0.9$ in the region $f_p\approx 5f_0=8.05~$Hz we will
find attractor P5, which is represented by stars in the case of
increasing frequency, and by a full line above them in the case of
decreasing frequency. In Fig. \ref{experimental}(a) all full lines
represent pendulum behavior that for a decreasing frequency go into
rotating motion, resulting in an abrupt increase in $|\omega|$. In 
Fig.\ref{experimental}(b) this is not always the case; for P5, at
$f_p=7.7~$Hz it ends up going back to period-1 oscillations. This
difference can be explained by the coexistence of periodic rotations
and oscillations. The coexistence of period-1, period-4 and period-5
oscillatory attractors in the small range $7.8<f_p<8.05$ Hz is
noteworthy. 

\section{Melnikov subharmonic function}
Parameter thresholds for the occurrence of resonant periodic
oscillations are derived by applying the Melnikov method for
subharmonic solutions. The planar simple pendulum is described by the
(Hamiltonian) vector field   
$f(\theta, \omega) = [\omega\mbox{\ \ }-\sin{(\theta)}]^\intercal$. 
The phase space has a pair of homoclinic orbits bi-asymptotic to the  
saddle points $(\theta, \omega) = (\pm \pi, 0)$, and their interior is
filled by a continuous family of periodic oscillations 
$q^{\alpha}(t) = (\theta^\alpha(t), \omega^\alpha(t))$, with $\alpha$
being a label, described by  
\begin{equation} \label{oscillations}
\begin{split}
\cos\left(\frac{\theta^\alpha}{2}\right) &= k\sn(t,k), \quad
\sin\left(\frac{\theta^\alpha}{2}\right) = \dn(t,k), \\
\omega^\alpha &= \dot{\theta}^\alpha = 2k\cn(t,k)
\end{split}
\end{equation}
where $\sn(t,k)$, $\cn(t,k)$ and $\dn(t,k)$ are the Jacobi elliptic
functions and $k \in [0,1]$ is the elliptic modulus
\cite{int-elipticas}. The period of these solutions is
$T_\alpha=4K(k)$, where $K(k)$ is the complete elliptic integral of
first kind. Now, perturb this system with the time-dependent and
periodic (of period $T_p=2\pi/\Omega_p$) vector field 
$g(\theta,\omega,P,\Omega_p,\beta)=[0 \mbox{\ \ }
  -\epsilon P\cos{(\Omega_p \tau)\sin{(\theta-\phi)}} -
  \beta\omega]^\intercal $,  
with $0<\epsilon \ll 1$, and consider the quantity
\begin{equation} \label{integral}
  M^{m}(t_0) = \int_{0}^{mT_p} - \omega^\alpha(t) \left[ P
    \cos{\Omega_p t} \sin{(\theta^\alpha-\phi)}-\beta\omega^\alpha
    \right]\ud t,   
\end{equation}
which  is a path integral computed along the periodic oscillation 
$q^\alpha(t)$ of period $T_\alpha = mT_p$, that is, a solution
satisfying the resonance condition 
\begin{equation} \label{resonance_condition}
  4K(k) = m \frac{2\pi}{\Omega_p}. 
\end{equation}
Theorem 4.6.2 of \cite{guck-holmes} states that if $M^m(t_0)$ has
simple zeros, meaning   $M^{\nicefrac{m}{n}}(t_0) = 0$, but
$\frac{\partial M^{\nicefrac{m}{n}}}{\partial t_0} (t_0) \neq 0$,
then the perturbed system has a solution that is a
periodic subharmonic oscillation of period $m T_p$, arising from a
saddle-node bifurcation.     

Substituting Eqs. \eqref{oscillations}, subject to the resonance
condition \eqref{resonance_condition}, into Eq. \eqref{integral}, we
have
\begin{eqnarray} \label{mel_integral}
  M^{m}(t_0) = - 4 k^2 \beta I_1 -4 k^2 P \cos \phi I_2 &+& \\ \nonumber
  2 k P \sin \phi I_3 &-& 4 k^3 P \sin \phi I_4 ,
\end{eqnarray}
where
\begin{eqnarray} \nonumber
I_1 &=& \int_0^{mT_p} \cn^2(t,k) \ud t, \\ \nonumber
I_2 &=& \int_0^{mT_p} \cos [\Omega_p (t+t_0)] \cn(t,k) \sn(t,k)
      \dn(t,k) \ud t, \\ \nonumber
I_3 &=& \int_0^{mT_p} \cos [\Omega_p (t+t_0)] \cn(t,k) \ud t, \\ \nonumber
I_4 &=& \int_0^{mT_p} \cos [\Omega_p (t+t_0)] \cn(t,k) \sn^2(t,k) \ud t. 
\end{eqnarray}
Imposing $M^{m}(t_0)$ has simple zeros we obtain the loci of the
saddle-node bifurcations generating a subharmonic resonant periodic
orbit, i.e., the minimum values $R^m$ such that for $P>R^m$ we might
observe stable oscillatory motion of period $m T_p$ in the parameter
space.

If $m$ is even, $I_3=I_4=0$, and $M^{m}(t_0)$ can only have simple
zeros if $I_2$, that comes from the vertical excitation component, is
non identically null. In this case, $R^m_{even}$, is given by   
\begin{equation} \label{melnikov_even}
  R^m_{even} (\Omega_p) = \frac{4 \beta [E(k) - k^{\prime 2} K(k)]}{\pi \cos
    \phi \Omega_p^2}  \sinh (\Omega_p K^\prime(k)),
\end{equation}
where $k^{\prime 2} = 1 - k^2$ is the complementary elliptic modulus
and $K^\prime(k) = K(k^\prime)$. 
On the other hand, if $m$ is odd, $I_2 = 0$, and $M^{m}(t_0)$ can only
have simple zeros if $I_3$ or/and $I_4$, which are both due to the
additional torque $\tau$ caused by the tilt in the pivot motion, are
non identically null. Then, $R^m_{odd}$ is given by  
{\small 
\begin{equation} \label{melnikov_odd}
  R^m_{odd} (\Omega_p) = \frac{4 \beta [E(k) - k^{\prime 2} K(k)]}
  {\pi \sin \phi \left\{ \left[ \frac{2E(k)}{K(k)}-1\right] 
    \sech(\Omega_p K^\prime(k)) + \frac{\pi^2 S}{K^2(k)}  \right\}}. 
\end{equation} }
where
{\small 
\begin{equation} \label{soma}
\begin{split}  
  S =   
  \sum_{l=0}^{\infty} \frac{a_{+}(l)}{2} 
  \csch \left[\frac{a_+(l)}{m} \Omega_p K^\prime(k) \right] 
  \sech \left[\frac{b(l)}{m} \Omega_p K^\prime(k) \right] + 
  \\ \nonumber
  + \sum_{\substack{l=0 \\ l\neq\frac{m-1}{2}}}^{\infty} \frac{a_{-}(l)}{2}
  \csch \left[\frac{a_{-}(l)}{m} \Omega_p K^\prime(k) \right]
  \sech \left[\frac{b(l)}{m} \Omega_p K^\prime(k) \right],
\end{split}
\end{equation}
}
with $a_{+}(l) = 2l+1+m$, $a_{-}(l) = 2l+1-m$ and $b(l) = 2l+1$. 

The elliptic modulus $k$ is determined from $\Omega_p$ through the
resonance condition, and it is worth noticing that for $\Omega_p>m$
there is no $k$ such that Eq. \eqref{resonance_condition} is satisfied. 
It is remarkable that for the vertically excited pendulum we have
$\sin{\phi}=0$, and no Melnikov condition for the existence of odd
oscillations can be determined, that is, $M^{m}(t_0)$ does not have
simple zeros in this case.

\begin{figure}
  \centering
  \includegraphics[width=8cm,height=5cm]{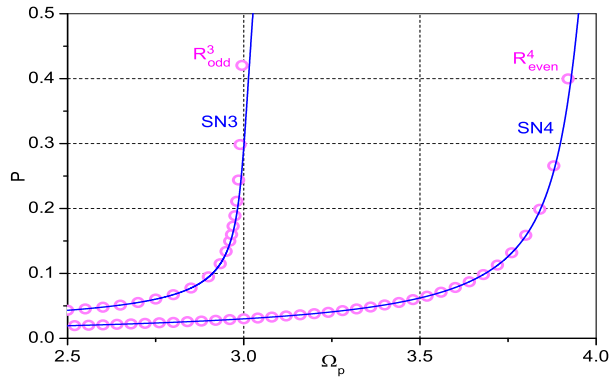}
  \caption{(Color online). A zoom into the parameter space of the
    tilted parametric pendulum, representing the red dashed rectangle
    region indicated in Fig. \ref{esp-par}. Melnikov thresholds
    $R^3_{odd}$ and $R^4_{even}$ (empty circles) agree finely with the
    borders of resonance regions SN3 and SN4 (full lines),
    respectively.} 
  \label{final}
\end{figure}

In Fig. \ref{final} we see remarkable agreement between numeric and
analytic predictions for the loci of the saddle-node bifurcations
where the oscillatory resonant stable periodic attractors P3 and P4
appear in the region $P<0.5$. Full lines SN3 and SN4 were obtained
with AUTO \cite{auto, auto-2} (see Fig. \ref{esp-par}), and circles
indicated by $R^3_{odd}$ and $R^4_{even}$ represent $\{R^3_{odd}
\times \Omega_p\}$ and $\{R^4_{even} \times \Omega_p\}$
respectively. Therefore, the minimal value of P analytically
calculated by the Melnikov method provides the parameter values for
the onset of these resonances.       

\section{Conclusions}
The novelty was to show that odd resonant oscillations are likely to
be observed in the tilted parametric pendulum, enlarging the spectra
of possible resonances in the system. The occurence of odd resonances
when excitation is along a tilted direction was explained by the
Melnikov subharmonic function. We not only demonstrated that the loci
of saddle-node bifurcations that is the threshold obtained by
application of this method excellently agrees with those computed
using numeric continuation technique, what implies that primary
resonances are indeed described by this method, but also showed that
according to this approach, whereas even resonances are due to the
vertical excitation component, odd resonances are a result of an extra
torque that appears only in the tilted case, as a consequence of the
symmetry breaking of the equations of motion.    

\section*{Acknowledgements}
This work was supported by the Brazilian agencies FAPESP and CNPq. MSB
also acknowledges  the  Engineering  and  Physical  Sciences Research
Council grant Ref. EP/I032606/1. GID thanks Felipe A. C.  Pereira for
fruitful discussions.


\begin{thebibliography}{99}
\bibitem{note-mathieu} Ignoring friction, in the small 
  amplitude regime the differential equation for the angle
  displacement $\theta$ of the pendulum arm is the Mathieu equation
  \cite{mathieu}, from which many properties are known.
\bibitem{note-periodo-n} For a system subject to periodic perturbation
  of period $T_p$, the term ``period-n'' indicates the solution is
  periodic with period equal to $n T_p$.
\bibitem{note-parametric} When the Melnikov method is applied to the
  vertically excited parametric pendulum, and also to other similar
  parametric systems, such as the parametric pendulum subject to
  non-harmonic excitation \cite{nonharmonic} and the pendulum with
  varying length \cite{belyakov}, integrating Eq. \eqref{integral},
  we see that the term that can lead to non simple zeros of the
  Melnikov integral is non identically null only for even
  oscillations. Thus, the parameter thresholds for the existence
  for odd oscillations cannot be computed, at least through this
  method.
\bibitem{note-pendulum-parameters} The natural frequency of
  oscillation $f_0$ was obtained by measuring the period of small
  oscillations, and the friction parameter $b$ was obtained by
  following the natural decay of small oscillations.
\end{thebibliography}
\end{document}